\documentclass[reprint,aps,preprintnumbers,amsmath,amssymb,nofootinbib,epsf,epsfig]
{revtex4-2}
\usepackage[utf8]{inputenc}
\usepackage{graphicx}
\usepackage{amsmath}
\usepackage{amssymb}
\usepackage{array}
\usepackage{float}
\usepackage{booktabs}
\usepackage{multirow}
\usepackage{color}
\usepackage{caption}
\usepackage{subcaption}
\usepackage{natbib}
\usepackage{pifont}
\usepackage[colorlinks, citecolor=cyan]{hyperref}
\usepackage{slashed}
\usepackage{appendix}
\usepackage{comment}

\begin{document}
\newcommand{\bea}{\begin{eqnarray}}
\newcommand{\eea}{\end{eqnarray}}
\newcommand{\ldm}{\Delta m_{31}^2}
\newcommand{\sdm}{\Delta m_{21}^2}
\newcommand{\dcp}{\delta_{CP}}
\newcommand{\be}{\begin{equation}}
\newcommand{\ee}{\end{equation}}
\newcommand{\nn}{\nonumber}
\newcommand{\mrm}[1]{\mathrm{#1}}
\newcommand{\mc}[1]{\mathcal{#1}}
\newcommand{\del}{\partial}
\newcommand{\dd}{\mathrm{d}}
\newcommand{\blue}[1]{{\color{blue} #1 }}
\global\long\def\d{\partial}
\def\ds{\displaystyle}
\def\s1{\hat s}
\def\para{\parallel}
\def\CP{{\it CP}~}
\def\cp{{\it CP}}
\def\ml{m_\mu}

\title{\large Enhancing DUNE Physics Sensitivity with Light and Charge Calorimetry}
\author{Jogesh Rout$^a$}
\email{jogesh.rout1@gmail.com}
\author{Suchismita Sahoo$^b$}
\email{suchismita@cuk.ac.in}
\affiliation{$^a$Department of Physics, Shree Ram College, Rampur, Subarnapur-767045, India \\ $^b$Department of Physics, Central University of Karnataka, Kalaburagi-585367, India}
\begin{abstract}
    We investigate the potential of light calorimetry in liquid argon time projection chambers and its intrinsic self compensation properties, emphasizing its advantages alongside conventional charge calorimetry. Previous studies have demonstrated that light calorimetry can achieve energy resolution comparable to advanced charge based techniques, particularly for GeV scale neutrinos. In this work, we explore the complementarity of light calorimetry with charge calorimetry for precision measurements of key physics parameters in the DUNE, including CP violation (CPV) and mass hierarchy determination. While charge calorimetry provides superior resolution in CP phase measurements, light calorimetry independently offers significant insights into CPV and mass hierarchy sensitivities. Furthermore, our exposure versus CPV sensitivity studies indicate that the $5\sigma$ discovery potential is reached faster using light and charge calorimetry than with the traditional TDR based reconstruction method. These findings highlight the promising role of light calorimetry as a simple yet effective reconstruction method, serving as a complementary approach to enhance the physics capabilities of DUNE.
\end{abstract}

\maketitle
\flushbottom

\section{Introduction}
\label{sec:intro}


Neutrino oscillations, the phenomenon in which neutrinos oscillate between different flavors as they travel through space at varying energies and distances, have been experimentally confirmed~\cite{PhysRevLett.81.1562, PhysRevLett.89.011301, PhysRevD.74.072003} and are now a cornerstone of neutrino physics. The parameters governing these oscillations, such as neutrino masses and mixing angles, have been measured with high precision, as demonstrated by recent global fits~\cite{Esteban:2024eli}. 
Despite significant advancements in measuring oscillation parameters, several unresolved questions remain within the framework of mass induced neutrino oscillations. Key open issues include determining the neutrino mass ordering (i.e., the sign of $\Delta m^{2}_{31}$), pinpointing the value of the CP violating phase ($\dcp$), and identifying the correct octant of the mixing angle $\theta_{23}$. Further improvements in the precision of these measurements are essential to advancing our understanding of neutrino physics. Resolving the neutrino mass ordering would provide deeper insights into the structure of the neutrino mass matrix and help distinguish between various theoretical models of neutrino masses~\cite{Albright:2006cw}. Additionally, determining the CP violating phase and the mass ordering plays a crucial role in the leptogenesis scenario, which seeks to explain the matter and antimatter asymmetry in the universe~\cite{Fukugita:1986hr}.

Accurate measurement of neutrino properties and interactions is a fundamental goal in modern particle physics. To advance neutrino oscillation experiments, such as probing CP violation  in the lepton sector and testing the three neutrino mixing framework, systematic uncertainties must be controlled at the percent level~\cite{DUNE:2020ypp,Hyper-Kamiokande:2018ofw, Coloma:2012ji}. 
One of the main challenges in these experiments is the precise reconstruction of neutrino energy, which depends on accurate interaction modeling, detector response, and energy reconstruction techniques. This is crucial for both cross-section measurements and oscillation studies as a function of baseline and neutrino energy. The reconstruction of neutrino energy requires a full kinematic reconstruction of all final state particles in a neutrino interaction, including both leptons and hadrons. Additionally, it must account for missing energy contributions from particles below detection thresholds, energy deposited in inactive materials, and neutral particles that escape detection \cite{Ankowski:2016jdd, ANDREOPOULOS201087, PhysRevD.93.051104}. 

Liquid argon has primarily been employed as a scintillation calorimeter in dark matter experiments such as DEAP-3600~\cite{Ajaj:2019imk}, WArP~\cite{Acciarri:2010it}, DarkSide-50~\cite{Agnes:2014bvk}, and DarkSide-20k~\cite{Aalseth:2017fik}. These experiments are optimized for detecting very low energy signals and emphasize the ability to distinguish between electronic and nuclear recoils. Consequently, significant effort has been dedicated to understanding the scintillation mechanism in liquid argon, including light yield, time structure of scintillation, response to different types of energy deposition, and optical transport properties. In contrast, the application of LAr scintillation for reconstructing the energies of hundreds of MeV or more higher energy particles has not been thoroughly investigated. Compared to other noble liquids such as liquid krypton~\cite{Fanti:2007vi} or liquid xenon~\cite{Baldini:2016tba}, liquid argon has a longer radiation length, which necessitates the use of larger detector volumes for effective calorimetric energy measurements at these scales. As a result, its adoption for high energy particle calorimetry remains relatively limited. 

Nevertheless, liquid argon has found a transformative application in neutrino physics through the development of the Liquid Argon Time Projection Chamber (LArTPC), a groundbreaking advancement in detection technology.  Originally proposed~\cite{Rubbia:1977zz} in 1974 by Nobel laureate Carlo Rubbia, it has played a key role in shaping modern neutrino experiments, beginning with the iconic ICARUS project~\cite{ICARUS:2004wqc}. Since then, many liquid argon based experiments~\cite{bagby2021icarus, Acciarri_2017, MicroBooNE:2015bmn, abud2022protodune,abi2020dunephysics} have been developed, with some currently operating at Fermilab and others, such as the DUNE experiment~\cite{DUNE:2020ypp}, being proposed for the future. When a high energy particle passes through liquid argon (LAr), it ionizes argon atoms, creating electron ion pairs ($e^{-}$ and $Ar^{+}$). At the same time, some argon atoms are excited to higher energy states ($Ar^{*}$) without ionization~\cite{doke2002absolute}. These excited atoms relax to their ground state, emitting scintillation light in the vacuum ultraviolet (VUV) range, 128 nm~\cite{DOKE199962}. Free electrons generated during ionization may recombine with positive ions, producing additional excited argon atoms. These excited atoms can interact with nearby argon atoms to form argon dimers. When the dimers return to their ground state, they emit VUV photons with a wavelength of 128 nm. If recombination occurs, the number of free electrons available for collection (charge signal) decreases, while the scintillation light (light signal) increases ~\cite{kubota1978recombination,hitachi1992luminescence}. Conversely, in the presence of a strong electric field, recombination is suppressed as the field pulls electrons away from ions, resulting in more charge collection and reduced light production. The balance between charge collection and scintillation light production depends strongly on the electric field. Recombination processes in LAr detectors are also closely tied to the ionization energy deposition per unit path length (dE/dx), which characterizes the local energy density deposited by a particle. At higher dE/dx, such as with slow, heavily ionizing particles (e.g., alpha particles or highly charged ions), ionization density is greater, increasing the probability of electron ion recombination~\cite{Segreto:2024xnp, PhysRevA.36.614}. This leads to reduced charge collection but enhanced scintillation light production. In contrast, at lower dE/dx, such as with minimally ionizing particles (e.g., high energy muons), ionization density is low, resulting in sparse distributions of electrons and ions. In such cases, recombination is less likely, leading to higher charge collection and lower light contribution. In regions with high ionization density (e.g., near the Brag peak of a particle or for low energy protons), recombination is more likely due to the dense packing of electrons and ions. This increases scintillation light production while decreasing charge collection, especially in low field regions. The anti-correlation between charge and light signals provides complementary information about the energy deposited by the particle and the nature of the interaction. In LArTPCs, the total energy deposited is calculated by combining measurements of both collected charge and scintillation light. Recombination reduces the charge signal while contributing to the light yield. By leveraging this anti-correlation, LArTPCs can achieve precise energy reconstruction, making them invaluable for understanding particle interactions. More details of energy dissipation mechanisms in LAr can be found in Refs.~\cite{doke1981fundamental,doke1990estimation,segreto2021properties,szydagis2021review}.

Building on these principles, liquid argon scintillation provides a complementary probe to charge collection in LArTPC detectors, helping to mitigate several sources of systematic uncertainty in neutrino energy reconstruction. First, recombination losses, where free electrons recombine with ions and are therefore not collected, are compensated by an increase in scintillation light, preserving the total detectable signal. Second, low-energy hadrons that fall below the charge detection threshold can still produce scintillation photons, allowing partial recovery of their energy deposits. Third, while final state interactions (FSIs) introduce model-dependent uncertainties, scintillation light offers a less biased observable, as it captures energy deposits regardless of the details of the final state. Fourth, detector non-uniformities that affect charge collection efficiency are partially mitigated by the isotropic nature of scintillation light production, which helps average out spatial variations. Finally, neutrons, which often escape without ionizing, can still interact and generate scintillation signals, enabling partial recovery of otherwise missing energy. Altogether, these properties make scintillation an essential component for improving calorimetric precision and reducing systematic uncertainties in liquid argon based neutrino detectors.

In the past, light based energy measurement was not widely used in LArTPC neutrino detectors~\cite{majumdar2021review}. This was mainly due to poor photon collection and strong variation in light response across the detector volume. However, the situation is expected to improve in future detector designs. One such example is the Aluminum Profiles with Embedded X-ARAPUCA (APEX)~\cite{Marinho:2025cyc} design planned for the DUNE Phase-II~\cite{DUNE:2024wvj, DUNE:2025gyl} far detectors. This design proposes to install large area photon detectors along the full field cage walls, increasing the optical coverage to nearly 60\%. Early simulation results show that this setup could achieve an average light yield of around 180 photoelectrons per MeV, which is comparable to that of established liquid scintillator detectors used in neutrino experiments \cite{kamland2003,borexino2014,dayabay2012,lsnd2001,miniboone2018} and other experiments~\cite{behrens1990zeus,acosta1991lead}. These developments make it a good time to evaluate the physics potential of using LArTPCs as light calorimeters and how this method compares to, and enhances, the more commonly used charge based calorimetry \cite{zhang2021calibration, chen2021tpc, ereditato2005lar}. 

The motivation for this study is to investigate the physics potential achievable using light information alone in the DUNE detector. Our objective is to demonstrate that even a relatively simple reconstruction based solely on scintillation light can yield physics results that are complementary to those obtained from traditional charge-based techniques. This is particularly compelling because light-based reconstruction is intrinsically less complex and more computationally efficient than charge reconstruction. By highlighting the complementary nature of light-based analyses and their potential for cross-validation with charge-based results, we aim to promote broader use of light information within the DUNE collaboration. In this study, we exploit the capabilities of the LArTPC as a dual calorimeter~\cite{DUNE:2017pqt}, utilizing both ionization charge and scintillation light to estimate neutrino energy. The light calorimetry in the LArTPC is self compensating~\cite{Ning:2024zxg}, as the energy deposited in liquid argon involves mechanisms of charge recombination and scintillation light production. We investigate the sensitivity of DUNE to leptonic CP violation, CP phase resolution, neutrino mass hierarchy, and the octant of $\theta_{23}$, using energy reconstruction methods that are discussed in detail in paper~\cite{Ning:2024zxg}. Furthermore, we compare different energy reconstruction approaches one based solely on charge measurements and the other on light measurements and evaluate their respective contributions to the improvement of physics sensitivity. 

The remainder of this paper is structured as follows: 
Section II details the experimental setup of the DUNE experiment, including simulation procedures. In Section III, we discuss the energy reconstruction methods based on both light and charge measurements. The main results, along with a qualitative discussion, are presented in Section IV. Finally, we conclude the paper in Section V.
\section{Experimental Setup and Simulation Details}
\subsection{Experimental Setup}

The DUNE experiment~\cite{DUNE:2020jqi} aims to study neutrinos over a 1,300 km baseline, from their production at Fermilab, Chicago to detection at the Sanford Underground Research Facility (SURF), South Dakota. To analyze the sensitivity of the experiment to the measurement of standard unknowns, we adopt four primary benchmark configurations for DUNE:  
\begin{itemize}
    \item \textbf{TDR}: The Technical Design Report outlines the full experimental design and methodology of the DUNE experiment~\cite{DUNE:2020jqi, DUNE:2021cuw}.
    
    \item \textbf{Q}: A charge reconstruction method introduced by A.~Friedland \textit{et al.}, offering an alternative framework for charge analysis~\cite{Friedland:2018vry}.

     \item \textbf{Q3}: A high resolution charge imaging technique developed to enhance energy reconstruction accuracy~\cite{Ning:2024zxg}.
    
    \item \textbf{L1}: A light calorimetry technique that sums the total detected light (L) from all particle interactions and scales it to estimate the incident neutrino energy~\cite{Ning:2024zxg}.
\end{itemize} 
Further details on these configurations are provided in Section~\ref{sec:res}.
  In all configurations, we assume a 40 kton LArTPC detector and a 120 GeV proton beam operating at 1.2 MW beam power, delivering 1.1 $\times$ 10$^{21}$ POT per year. We assume equal runtime for neutrino and antineutrino modes, expressed in calendar years (CY). The term CY is used to differentiate from DUNE's actual runtime schedule, which accounts for an uptime of 57$\%$ per calendar year.  Further details on systematic errors and efficiencies are available in Refs.~\cite{DUNE:2021cuw}. 
\subsection{Simulation Details}
\label{sec:sim}

To quantify the statistical sensitivity of the numerical simulations conducted in this study, we utilize the built-in $\chi^2$ function in the General Long-Baseline Experiment Simulator (GLoBES) software~\cite{Huber:2004ka, Huber:2007igb}:
\begin{equation}
\chi^2 = \min_{\xi_s, \xi_b} \left[2 \sum_{i=1}^{n} \left( \tilde{y}_i - x_i - x_i \ln \frac{\tilde{y}_i}{x_i} \right) + \xi_s^2 + \xi_b^2 \right] \,.
\label{eqn:chisqdef}
\end{equation}
Here, $n$ denotes the total number of energy bins. The quantity $x_i = N_i^{\text{ex}} + N^b_i$ represents the total observed events in the $i$-th bin, consisting of the observed signal $N_i^{\text{ex}}$ and background $N^b_i$. The predicted number of events in each bin, including systematic uncertainties, is given by
\begin{equation}
\tilde{y}_i(\{\omega\}, \{\xi_s, \xi_b\}) = N^{\text{th}}_i(\{\omega\})[1 + \pi^s \xi_s] + N^b_i(\{\omega\})[1 + \pi^b \xi_b] \,,
\label{eqn:chisqerrors}
\end{equation}
where $N^{\text{th}}_i(\{\omega\})$ and $N^b_i(\{\omega\})$ are the expected signal and background events in the $i$-th bin, evaluated for a set of oscillation parameters, $\omega$. The charged current background depends on $\omega$, while the neutral current background is assumed to be independent of it. The parameters $\pi^s$ and $\pi^b$ denote the fractional systematic uncertainties on signal and background, respectively. Their associated pull variables, $\xi_s$ and $\xi_b$, are nuisance parameters accounting for deviations arising from these uncertainties.

We assume a constant line-averaged Earth matter density of $\rho = 2.848~\text{g/cm}^3$, based on the Preliminary Reference Earth Model (PREM)~\cite{Stacey_Davis_2008, DZIEWONSKI1981297}, and marginalize over a 2\% uncertainty in this density during the $\chi^2$ minimization. The normalization uncertainties are taken to be 5\% for muon neutrino signals and 2\% for electron neutrino signals. A common 5\% normalization uncertainty is applied to muon and electron neutrino backgrounds. Additionally, we include a 10\% uncertainty on the neutral current background and a 20\% uncertainty on the tau neutrino background~\cite{DUNE:2021cuw}.

Unless stated otherwise, we use the benchmark values for the standard neutrino oscillation parameters, $\theta_{12}$, $\theta_{13}$, $\theta_{23}$, and $\delta_{\text{CP}}$, which define the PMNS mixing matrix~\cite{ParticleDataGroup:2020ssz}. These values are listed in Table~\ref{tab1}. We marginalize over these parameters when necessary to obtain the minimum $\chi^2$.
\begin{table*}[htb]
\centering
\begin{tabular}{| c | c | c | c |}
\hline
Parameter & Best-fit value & 3$\sigma$ interval & Relative uncertainty \\
          & NO (IO)        & NO (IO)            & NO (IO) \\
\hline
$\theta_{12}$ ($^\circ$)        & 33.68               & 31.63 $\to$ 35.95             & 2.1\% \\
$\theta_{13}$ ($^\circ$)        & 8.56 (8.59)         & 8.19 $\to$ 8.89 (8.25 $\to$ 8.93) & 1.4\% (1.3\%) \\
$\theta_{23}$ ($^\circ$)        & 43.3 (47.9)         & 41.3 $\to$ 49.9 (41.5 $\to$ 49.8) & 3.1\% \\
$\Delta m_{21}^2$ ($10^{-5}~\text{eV}^2$) & 7.49    & 6.92 $\to$ 8.05              & 2.5\% \\
$\Delta m_{3\ell}^2$ ($10^{-3}~\text{eV}^2$) & 2.513 (--2.484) & 2.451 $\to$ 2.578 (--2.547 $\to$ --2.421) & 0.8\% \\
$\delta_{\text{CP}}$ ($^\circ$) & --148 (--86) & [--180, 180] & -- \\
\hline
\end{tabular}
\caption{Benchmark values of the three-flavor neutrino oscillation parameters, including best-fit values, 3$\sigma$ ranges, and relative uncertainties. If the $3\sigma$ upper and lower bounds of a parameter are $x_u$ and $x_l$, the 1$\sigma$ relative uncertainty is computed as $(x_u - x_l)/[3(x_u + x_l)]$~\cite{Esteban:2024eli}. Values for inverted ordering (IO) are shown in parentheses.}
\label{tab1}
\end{table*}
These benchmark values are consistent with the latest global fit results~\cite{Esteban:2024eli}.%
\footnote{For Majorana neutrinos, two additional phases appear in the PMNS matrix, though they do not affect neutrino oscillation probabilities.}

To evaluate the full sensitivity of the experiment, we compute the total $\chi^2$ as the sum of contributions from the four oscillation channels accessible at DUNE: muon neutrino disappearance $(\nu_\mu \rightarrow \nu_\mu)$ and electron neutrino appearance $(\nu_\mu \rightarrow \nu_e)$, both in neutrino and antineutrino modes:
\begin{equation}
\chi^2_{\text{total}} = \chi^2_{\nu_\mu \rightarrow \nu_e} + \chi^2_{\nu_\mu \rightarrow \nu_\mu}
+ \chi^2_{\bar{\nu}_\mu \rightarrow \bar{\nu}_e} + \chi^2_{\bar{\nu}_\mu \rightarrow \bar{\nu}_\mu} \,.
\label{eq:total-chisq}
\end{equation}

\section{Energy Reconstruction Methods}
\label{sec:res}
In LArTPC based neutrino detectors, the calorimetric reconstruction of neutrino energy is commonly employed to estimate the energy of incoming neutrinos interacting with LAr atoms. LArTPC operates as a dual calorimeter by capturing both charge and light signals, which serve as inputs for estimating energy. However, this reconstruction is subject to various inaccuracies due to several factors. These include nuclear effects in neutrino interactions, undetected energy carried away by secondary neutrinos, particles exiting the active detector volume, quenching of liquid argon ionization or excitation due to nuclear breakups, electronic noise in the charge readout system, electron attachment along the drift path, and electron-ion recombination effects.

In this work, we adopt the reconstruction methods (Q3 and L1) proposed by X. Ning \textit{et al.} \cite{Ning:2024zxg}. Their approach emphasizes the simplicity of Light Calorimetry (L1), which achieves performance comparable to more advanced charge imaging techniques, such as Q3, while avoiding the complexities associated with charge reconstruction. In the L1 method, the total detected scintillation light ($L$) from all particle activities is summed and scaled to estimate the incident neutrino energy, as described in Ref.~\cite{Ning:2024zxg}:
\begin{equation} E^{L1}_{\mathrm{rec}} = \frac{L}{0.42}. \end{equation}
In the Q3 approach, it assumes that electron ($e$) and hadron ($h$) charge activities can be effectively grouped using pattern recognition algorithms applied to 3D images (cf.\ Appendix~C from Ref.~\cite{Ning:2024zxg}). The charge contributions from $e$ and $h$ are scaled separately before being summed to estimate the incident neutrino energy. Furthermore, tracks longer than 2~cm are assumed to be reconstructible, with the $\mathrm{d}E/\mathrm{d}x$ along these tracks measurable. The energy deposited ($E_{\mathrm{dep}}$) along such tracks can be accurately reconstructed by applying corrections for charge recombination factors as outlined in Eq.~(17) of Ref.~\cite{Ning:2024zxg}. Additionally, we compare this charge reconstruction method with an alternative technique proposed by A.~Friedland \textit{et al.}~\cite{Friedland:2018vry}, where the author investigates the optimal scenario for incoming neutrino energy reconstruction, assuming the detector can identify all particles in an event and account for the small but frequent energy deposits from recoiled neutrons. The study concludes that this approach could improve reconstruction performance by up to a factor of 3 compared to the values reported in the TDR. Previous studies on physics sensitivity at DUNE have explored improved energy reconstruction using charge calorimetry alone~\cite{DeRomeri:2016qwo, Rout:2020emr, Chatterjee:2021wac}. In this work, we explore various reconstruction methods, including charge-based reconstruction (Q3) and light-based reconstruction (L1), both introduced in Ref.~\cite{Ning:2024zxg}, as well as an alternative charge-based reconstruction (Q) method from Ref.~\cite{Friedland:2018vry}. While Q3 and Q rely on detailed charge reconstruction and complex pattern recognition, the L1 method offers a much simpler alternative by using only the total scintillation light to estimate energy. Despite its simplicity, prior studies~\cite{Ning:2024zxg} suggest that L1 can achieve energy resolution comparable to charge-based techniques, making it a promising candidate for complementary or alternative reconstruction. This motivates our investigation into how light-only methods can contribute to DUNE's broader physics program. Our study aims to evaluate these approaches, with a particular focus on investigating light calorimetry for the first time.

The aforementioned energy reconstruction methods, except for the TDR, have been parameterized using the energy resolution function,  
\begin{equation}
R(E, E_r) = \frac{e^{-(E - E_r)^2 / 2\sigma^2}}{\sigma \sqrt{2\pi}} \, ,
\end{equation}  
where $E$ represents the true neutrino energy, $E_r$ is the reconstructed energy, and $\sigma$ denotes the energy resolution. The resolution $\sigma$ is defined as:  
\begin{equation}
\label{energy_res}
\sigma(E)/{\rm GeV} = a  (E/{\rm GeV}) + b  \sqrt{E/{\rm GeV}} + c \, ,
\end{equation}  
where $a$, $b$, and $c$ are the fit parameters. We assume the same energy resolution for neutrino and antineutrino modes, which is applicable for appearance and disappearance channels. Additionally, the energy resolution migration matrices for neutral current backgrounds, $\nu_e$ contamination, misidentified muons, and $\nu_\mu \to \nu_\tau$ backgrounds have been consistently applied for all cases, as provided in TDR~\cite{DUNE:2021cuw}.

\begin{figure}[htb]
    \centering
    \includegraphics[width=0.8\linewidth]{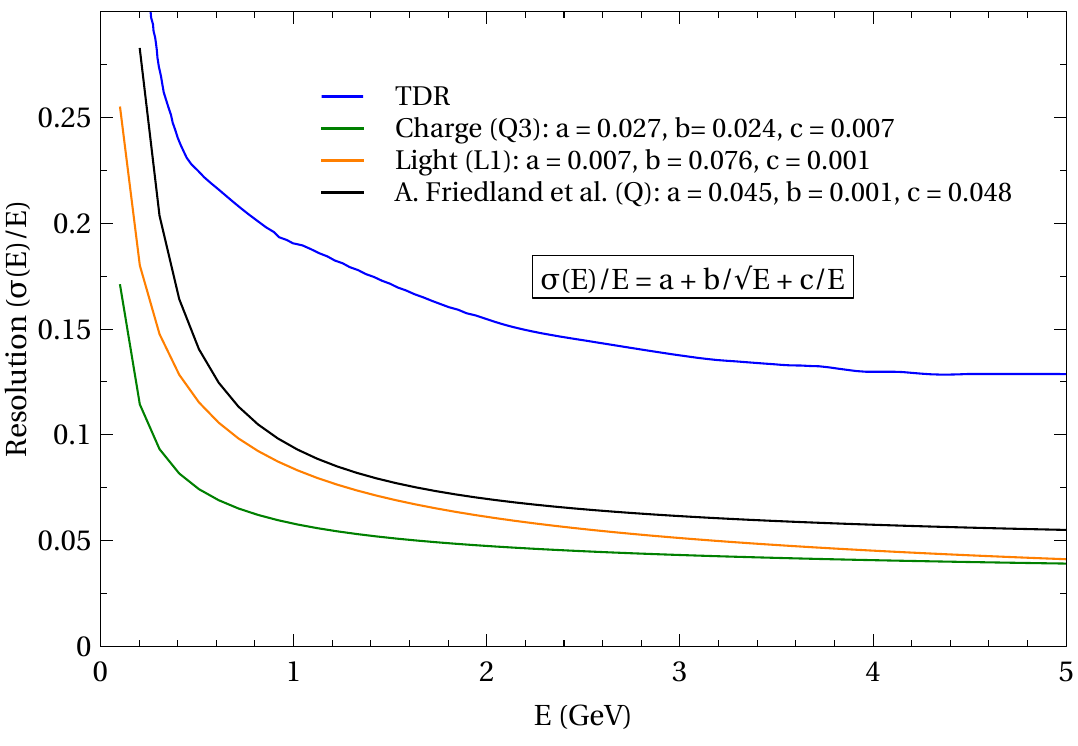}
    \caption{Energy resolution as a function of true neutrino energy in neutrino mode. The TDR (blue), Light Calorimetry (L1, green), Charge Calorimetry (Q3, orange), and the charge based method by A. Friedland et al. (Q, black) are shown. Fit parameters for each method are provided for reference.}
    \label{fig:energy-res}
\end{figure}
Fig.~\ref{fig:energy-res} shows the energy resolution as a function of the true energy of neutrinos in neutrino mode for TDR, charge calorimetry (Q3), light calorimetry (L1), and the charge based method by A. Friedland et al. (Q), represented in blue, green, orange, and black, respectively. The resolution reported in the TDR is the least precise among all methods, while charge calorimetry (Q3) demonstrates the highest resolution. Light calorimetry (L1) also exhibits improved resolution compared to the TDR and performs comparably to the method proposed by A. Friedland et al. (Q). Notably, for higher neutrino energy values, the resolution of light calorimetry closely approaches that of charge calorimetry.

\section{Results and Discussion}
\label{results}
\subsection{Sensitivity to CP Violation}
\label{cpv}


DUNE aims to determine whether CP violation occurs in the leptonic sector within the framework of the standard three flavor neutrino model.
The experiment's sensitivity to discovering CP violation at a given true value of $\delta_{CP}$ is evaluated by minimizing $\Delta\chi^2$, defined as 
\begin{equation}
    \Delta\chi^2_{CPV} = \min \big[ \Delta\chi^2_{CP}(\delta_{CP}^{\text{test}} = 0), \Delta\chi^2_{CP}(\delta_{CP}^{\text{test}} = \pi) \big]
\end{equation}
where $\Delta\chi^2_{CP} = \chi^2_{\delta_{CP}^{\text{test}}} - \chi^2_{\delta_{CP}^{\text{true}}}$. This leads to a distinct double peak structure in the sensitivity curve, centered around the CP violating phases, $\delta_{CP} = \pm\pi/2$~\cite{DUNE:2020jqi}.

\begin{figure}[hbt!]
    \centering
    \includegraphics[width=0.65\linewidth]{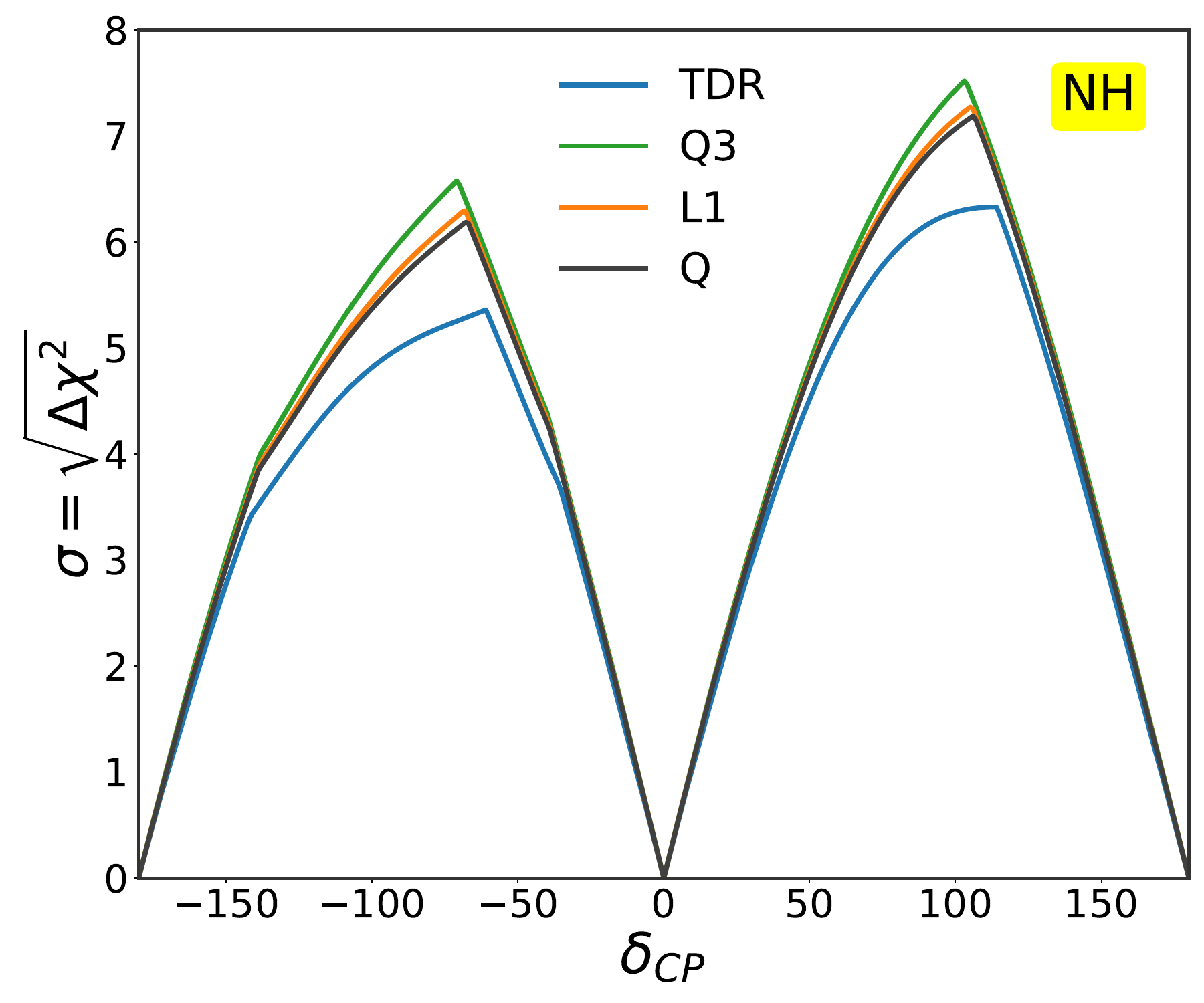}\quad
    \includegraphics[width=0.65\linewidth]{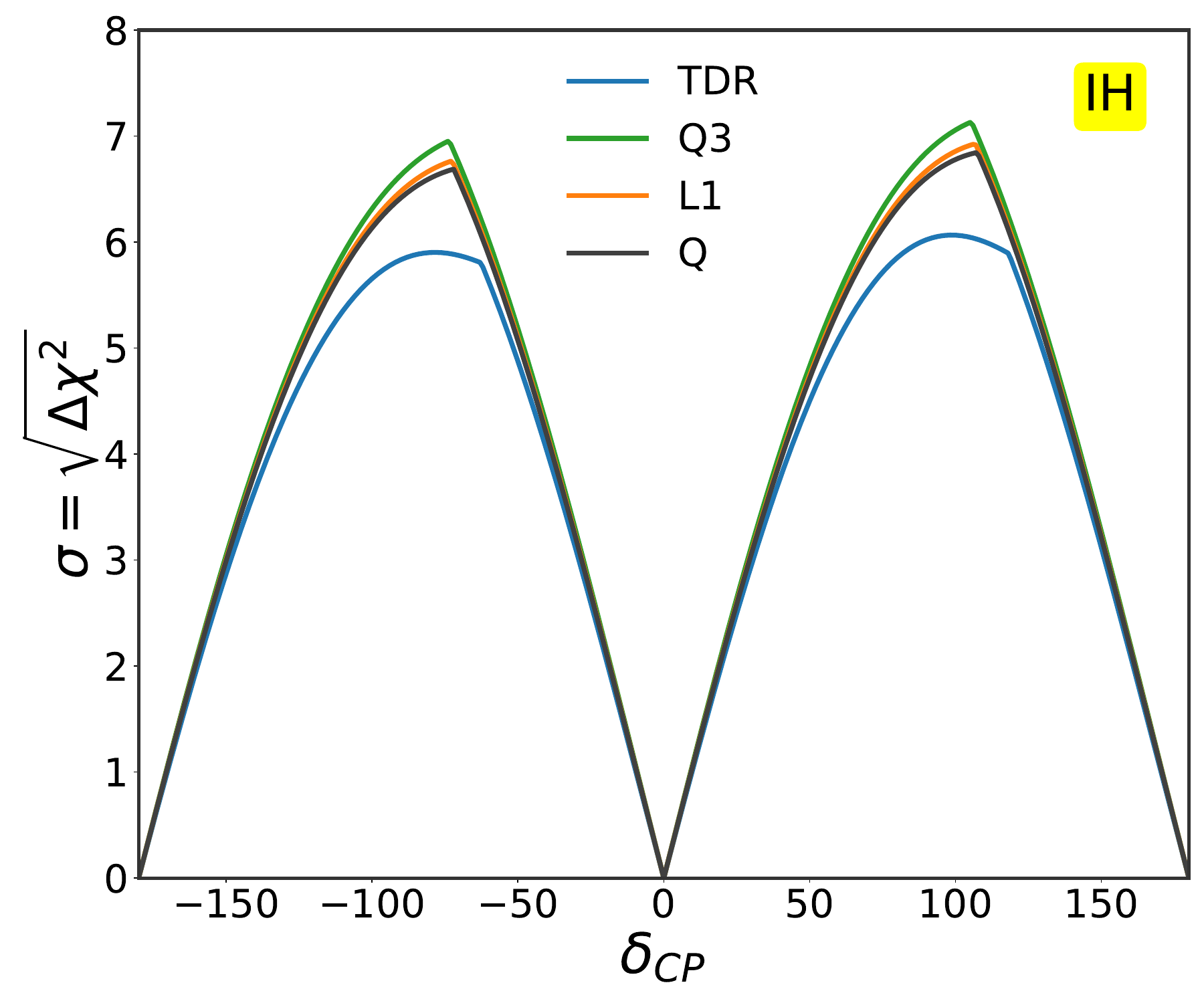}
    \caption{CP violation sensitivity as a function of the true value of $\delta_{CP}$ for different energy reconstruction methods. The left panel corresponds to the Normal Hierarchy (NH), while the right panel represents the Inverted Hierarchy (IH). The blue, black , green, and orange curves depict the sensitivity for the TDR, the charge-based energy resolution from A. Friedland et al. (Q), Charge Calorimetry (Q3), and  Light Calorimetry (L1) respectively.}
    \label{fig:cp-sens}
\end{figure}

Figure~\ref{fig:cp-sens} shows the significance of CP violation as a function of the true value of $\delta_{CP}$, assuming a total runtime of seven years (3.5 years in $\nu$-mode + 3.5 years in $\bar{\nu}$-mode) for four scenarios: TDR, Q, Q3, and L1. The left panel corresponds to the normal hierarchy, while the right panel corresponds to the inverted hierarchy. These conventions remain the same for all subsequent figures, except for exposure vs. sensitivity plots, where the runtime is fixed.

Among the four scenarios, Q3 demonstrates the highest sensitivity near the maximal CP violating phase values, $\delta_{CP} = \pm\pi/2$, while TDR exhibits the lowest sensitivity for both NH and IH. The L1 and Q methods yield nearly identical results for all values of $\delta_{CP}$. A sensitivity of $5\sigma$ can be attained near the maximal CP violating phase values for all four scenarios and both mass orderings. For IH, the peaks exhibit a very similar shape at $\delta_{CP} = -\pi/2$ and $\delta_{CP} = \pi/2$. The peaks are more pronounced for NH than IH at $\delta_{CP} = \pi/2$, whereas they are stronger for IH than NH near $\delta_{CP} = -\pi/2$. This figure illustrates the CPV sensitivity for a fixed exposure of  $336$~Kt-MW-CY  ($40$~Kt~$\times$~$1.2$~MW~$\times$~$7$~CY\footnote{CY represents the calendar year, which we distinguish from the actual runtime of DUNE, as its uptime is $57\%$ of CY. Therefore, DUNE must run longer to achieve the desired exposure.}), which may reveal additional features when varying the exposure.

\begin{figure}[hbt!]
    \centering
    \includegraphics[width=0.7\linewidth]{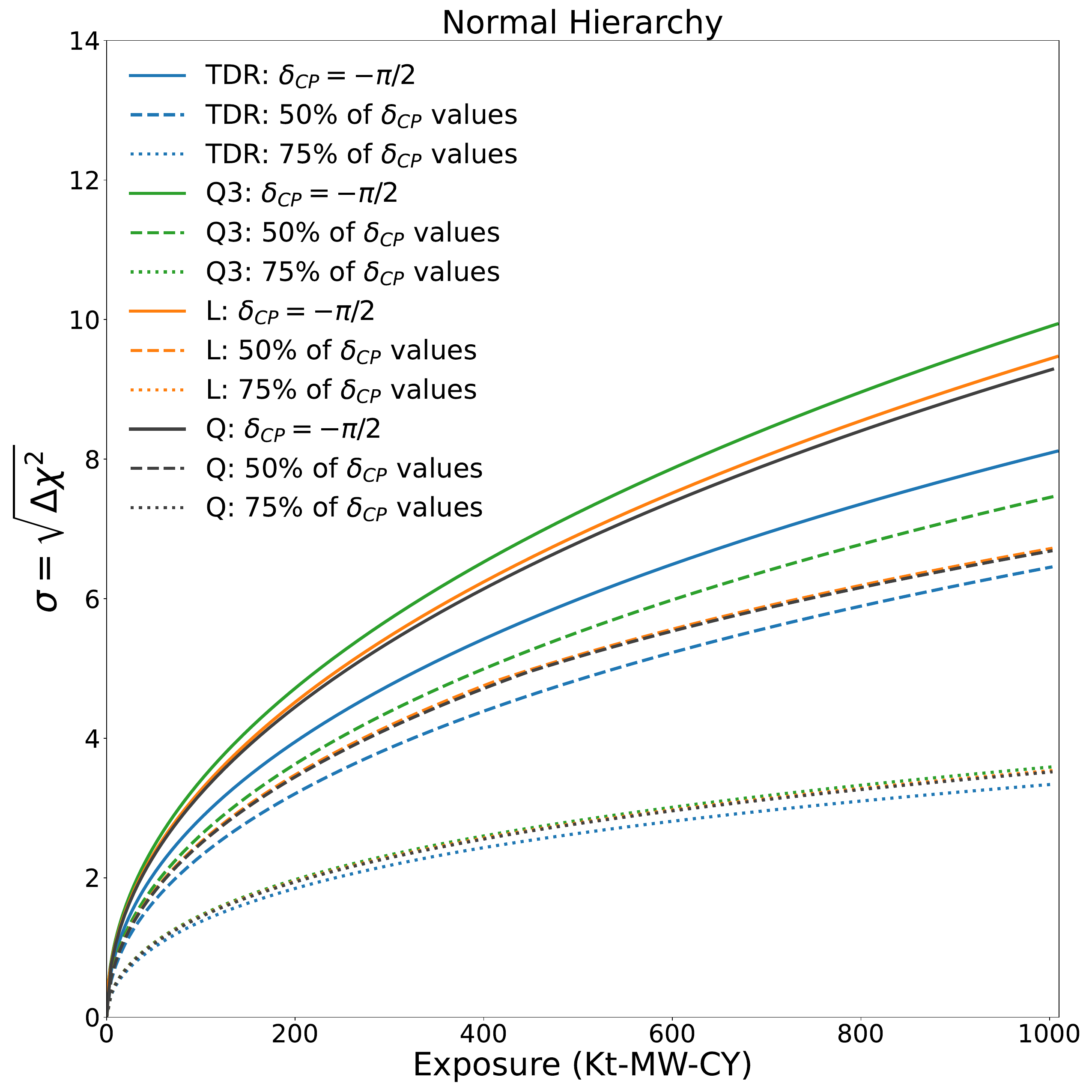}\quad
    \includegraphics[width=0.7\linewidth]{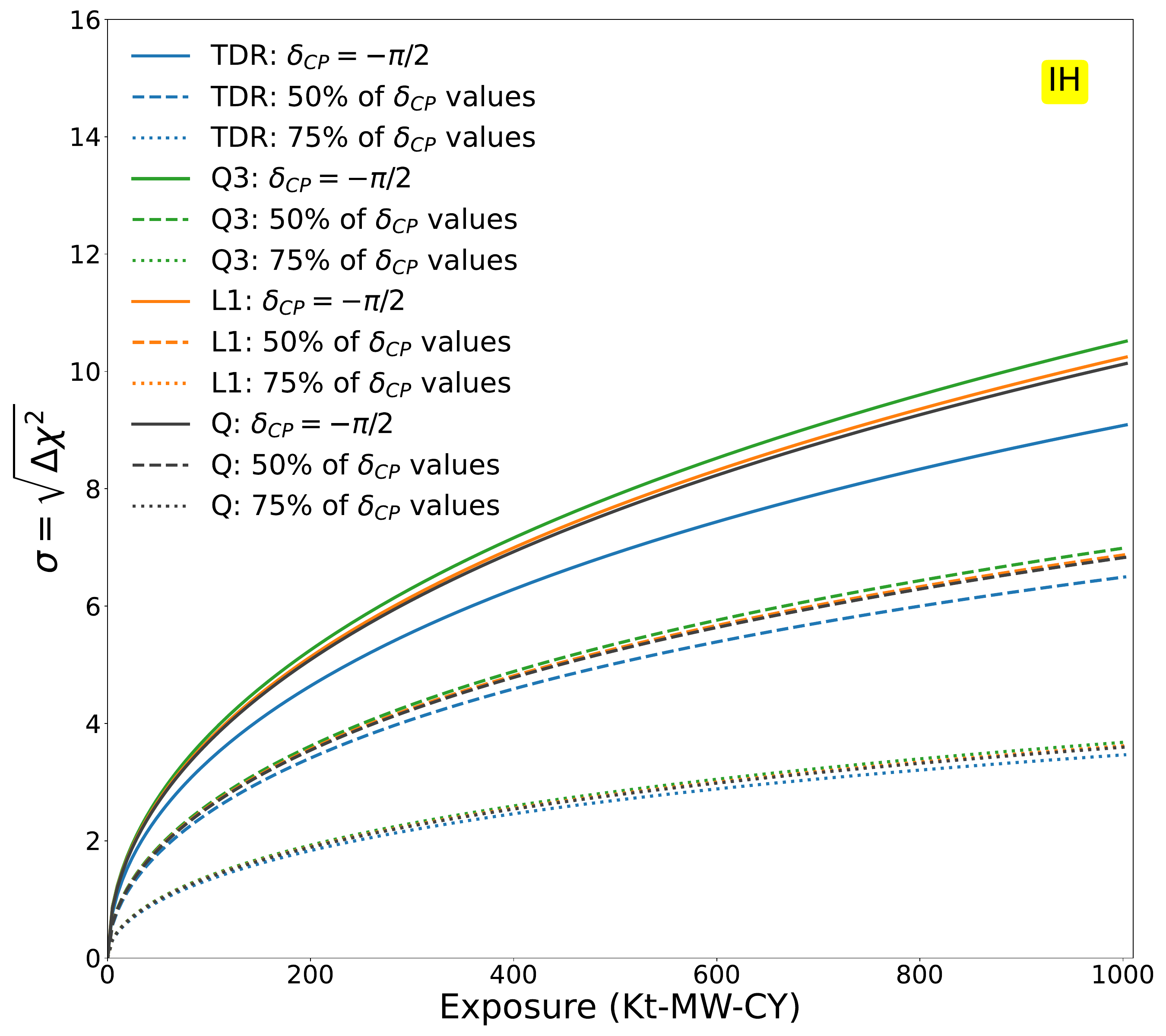}
    \caption{Same as Fig. \ref{fig:cp-sens} for CP violation sensitivity with exposure in Kt-MW-CY. Here, the solid curves represent sensitivity for $\dcp=-\pi/2$, while the dashed and dotted curves correspond to $50\%$ and $75\%$ of possible true $\delta_{CP}$ values, respectively. }
    \label{fig:cpsens-exp}
\end{figure}
Figure~\ref{fig:cpsens-exp} presents the significance level for CPV determination with exposure, measured in Kt-MW-CY  for all four scenarios and both mass orderings. The dashed and dotted lines indicate the sensitivity for 50\% and 75\% of  $\dcp$ values, respectively, while the solid curve corresponds to $\dcp=-\pi/2$. This convention is maintained for all subsequent exposure versus sensitivity figures. Assuming $\dcp=-\pi/2$, a $5\sigma$ sensitivity is achievable with NH after approximately $7$ years using TDR, $5.3$ years using Q, $5.2$ years using L1, and $4.7$ years using Q3 based reconstruction. In the case of IH, the required exposure is reduced to $4.9$ years for TDR, $4$ years for Q, $3.9$ years for L1, and $3.8$ years for Q3. For $75\%$ of $\dcp$ values, a $3\sigma$ sensitivity is attained after approximately $15$ years with TDR, $12.8$ years with L1, $13$ years with Q, and $12.4$ years with Q3 under NH, while IH requires slightly shorter exposures of $13.9, 12.5, 12.7$, and $12$ years, respectively. For $50\%$ of $\dcp$ values, a $5\sigma$ significance is reached after approximately $11.2 ~(10.3)$ years using TDR, $9.5 ~(9.3)$ years using Q, $9.4~ (9.1)$ years using L1, and $9.4~ (8.8)$ years using Q3 under NH (IH). Among the three alternative calorimetry methods Q3, L1, and Q, similar performance is observed for $75\%$ of true $\dcp$ values in both mass orderings. The overall comparison demonstrates that these three approaches outperform TDR, with L1 exhibiting a complementary performances with Q and Q3.


\subsection{CP Phase Resolution}
\begin{figure}[hbt!]
    \centering
    \includegraphics[width=0.65\linewidth]{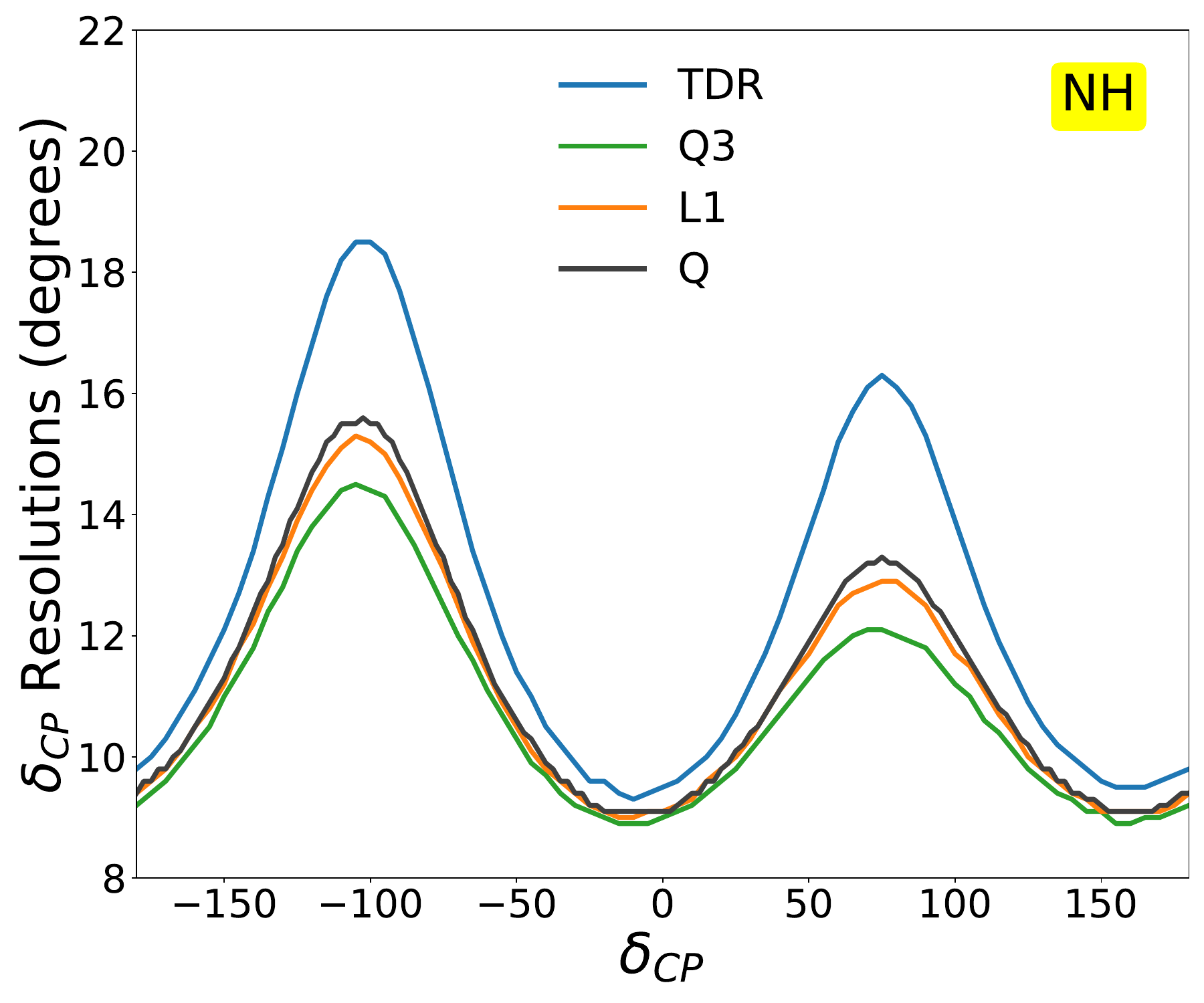} \quad
    \includegraphics[width=0.65\linewidth]{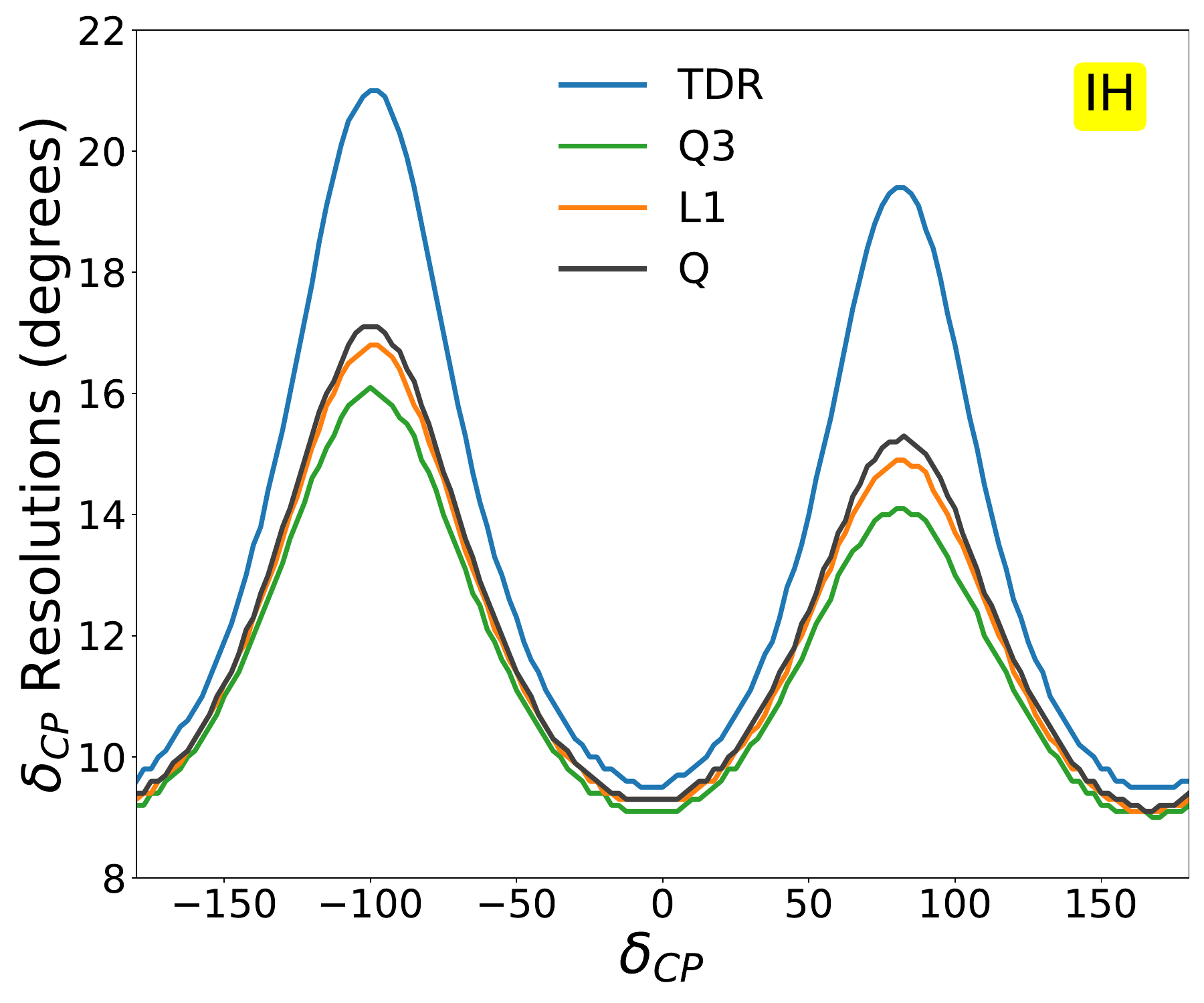}
    \caption{ $\delta_{CP}$ resolution (degree) as a function of true values of $\delta_{CP}$.  }
    \label{fig:cp-res}
\end{figure}

Given the strong potential to discover the violation of CP in the leptonic sector, precisely measuring the Dirac CP phase, $\dcp$, is expected to be a key objective of future neutrino oscillation experiments. We examine the impact of various calorimetric energy reconstruction methods on the precision of $\dcp$ measurements at DUNE. Assuming a total runtime of seven years (3.5 years in $\nu$-mode and 3.5 years in $\bar{\nu}$-mode), figure \ref{fig:cp-res} presents the variation in $\delta_{CP}$ resolution (in degrees) for different true $\delta_{CP}$ values across four scenarios: TDR, Q, Q3, and L1. 

\begin{figure}[hbt!]
    \centering
    \includegraphics[width=0.65\linewidth]{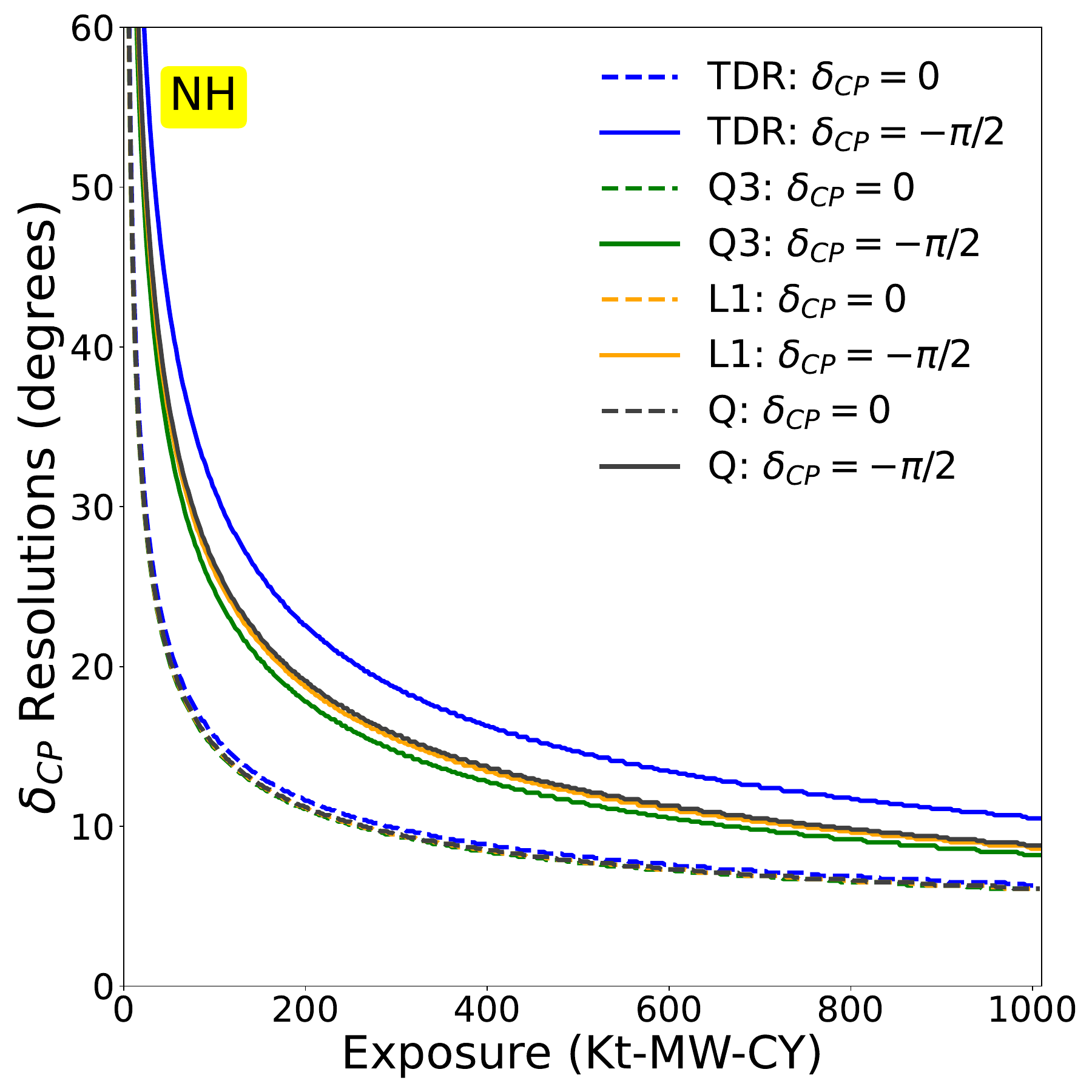}\quad
    \includegraphics[width=0.65\linewidth]{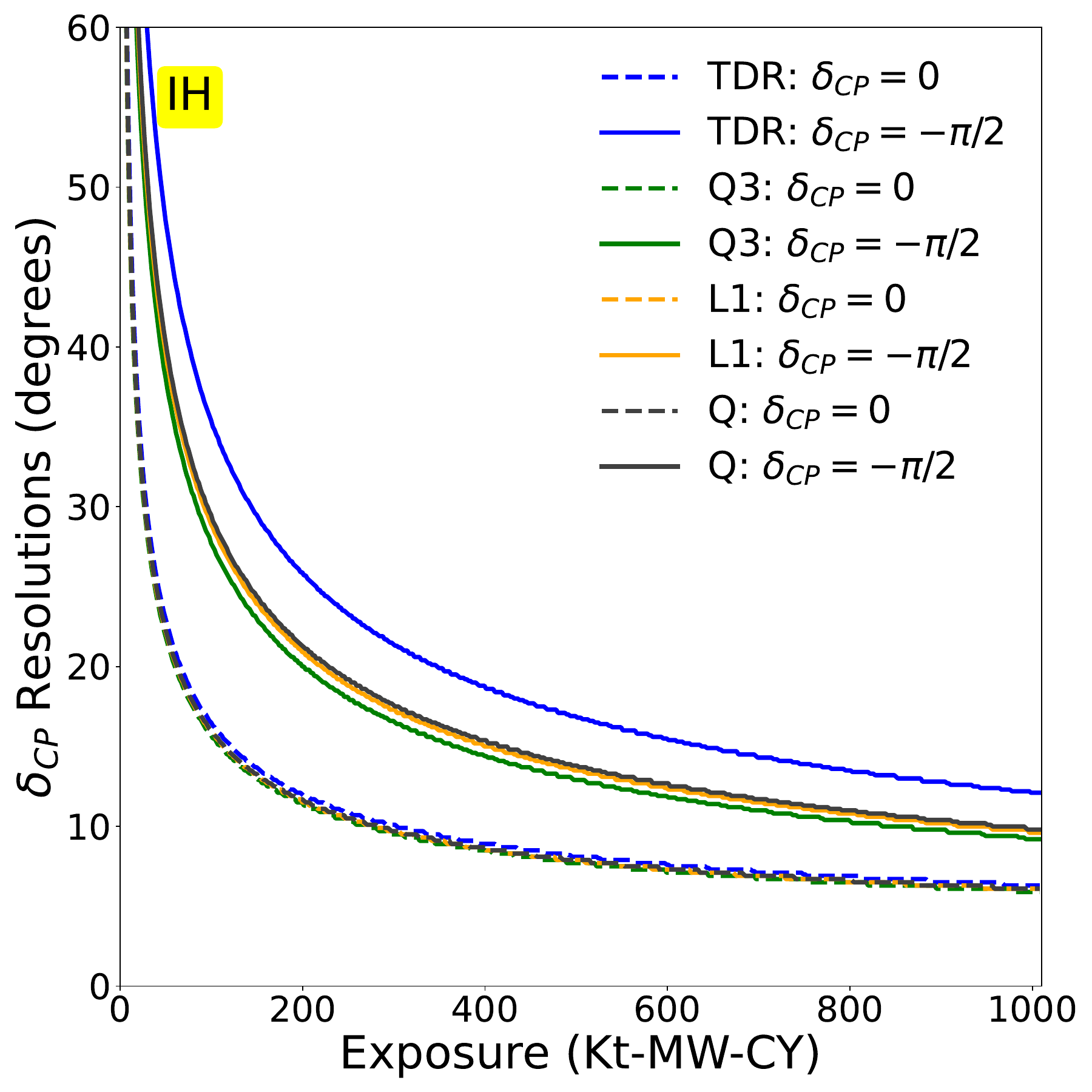}
    \caption{The resolution of $\delta_{CP}$ as a function of exposure (Kt-MW-CY) is shown for $\delta_{CP} = 0$ (dashed) and $\delta_{CP} = -\pi/2$ (solid).  }
    \label{fig:cpres-exp}
\end{figure}

The resolution is notably better near CP conserving values ($0, \pi$) compared to maximally CP violating values ($\pm\pi/2$) across all scenarios. Among these, TDR shows the poorest resolution around maximally CP violating values for both mass orderings, with the normal hierarchy generally yielding better resolution than the inverted hierarchy. Near CP conserving phases, the resolutions for Q, Q3, and L1 are nearly identical, but at maximally CP violating values, Q3 demonstrates the best resolution. At $\dcp=0$, the resolution is $9^\circ$ ($9.1^\circ$) for Q3, $9.1^\circ$ ($9.3^\circ$) for L1, $9.1^\circ$ ($9.3^\circ$) for Q, and $9.5^\circ$ ($9.5^\circ$) for TDR under NH (IH). Similarly, at $\dcp=\pm180^\circ$, the resolution is $9.2^\circ$ ($9.2^\circ$) for Q3, $9.4^\circ$ ($9.3^\circ$) for L1, $9.4^\circ$ ($9.4^\circ$) for Q, and $9.8^\circ$ ($9.6^\circ$) for TDR. The results indicate that both mass orderings achieve nearly equal resolutions near CP conserving phases, while the inverted hierarchy (IH) shows the poorest resolution at maximally CP violating phases. Additionally, Q and L1 provide almost identical resolutions near CP-conserving values. Importantly, all three alternative calorimetry methods (Q, Q3, and L1) consistently outperform TDR in terms of resolution for both hierarchies.
Figure~\ref{fig:cpres-exp} shows the $\delta_{CP}$ resolution for different exposures in Kt-MW-CY for TDR, L1, Q, and Q3, with true $\delta_{CP}$ values set to 0 and $-\pi/2$ for both NH and IH. 
Here, the solid lines represent $\dcp=-\pi/2$, while dashed lines correspond to $\dcp=0$. These values are selected because the resolution is generally worst at $\delta_{CP} = 0$ and best at $\delta_{CP} = -\pi/2$ among all possible $\delta_{CP}$ values.  It can be observed that the resolution for $\delta_{CP} = 0$ is consistently better than for $\delta_{CP} = -\pi/2$. With increasing exposure, all four scenarios yield nearly identical resolutions for $\dcp=0$, whereas for $\dcp=-\pi/2$, Q3 outperforms L1, Q, and TDR. L1 and Q provide nearly identical resolutions, except near $\dcp=\pm\pi/2$, in both NH and IH. For an exposure of $20$ years, the resolution at $\delta_{CP} = 0$ in NH (IH) is found to be $6.5^\circ$ ($6.5^\circ$) for TDR, $6.2^\circ$ ($6.1^\circ$) for L1, $6.2^\circ$ ($6.1^\circ$) for Q, and $6.1^\circ$ ($6.1^\circ$) for Q3. A resolution of $10^\circ$ at $\delta_{CP} = 0$ can be achieved after approximately $6 ~(6.3)$ years for TDR, $5.5 ~(5.9)$ years for Q, $5.4 ~(5.8)$ years for L1, and $5.3~ (5.6)$ years for Q3 in NH (IH). TDR consistently exhibits the worst resolution among the four scenarios, regardless of the mass ordering.

\subsection{Sensitivity to Mass Ordering}
\begin{figure}[hbt!]
    \centering
    \includegraphics[width=0.65\linewidth]{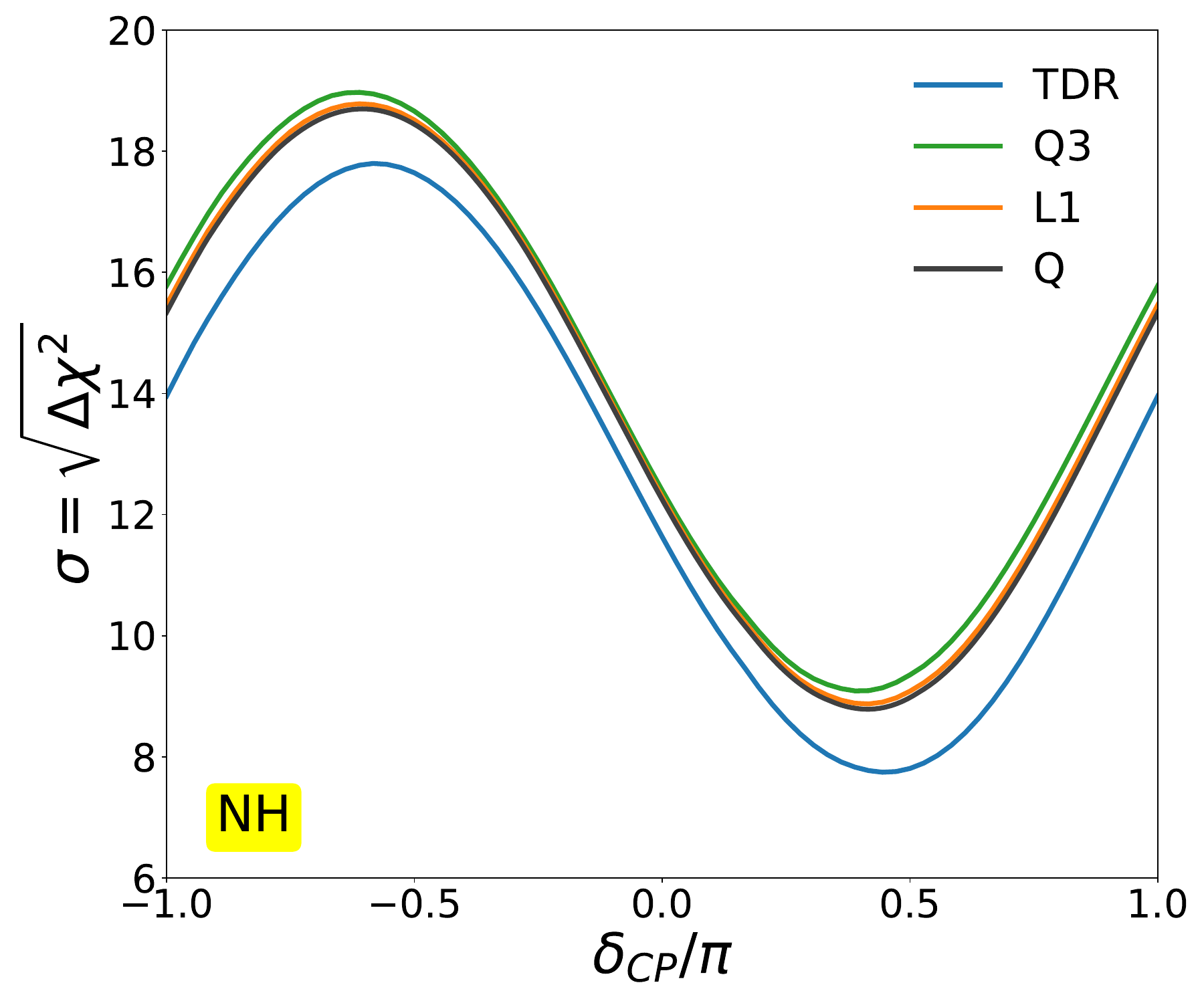} \quad
    \includegraphics[width=0.65\linewidth]{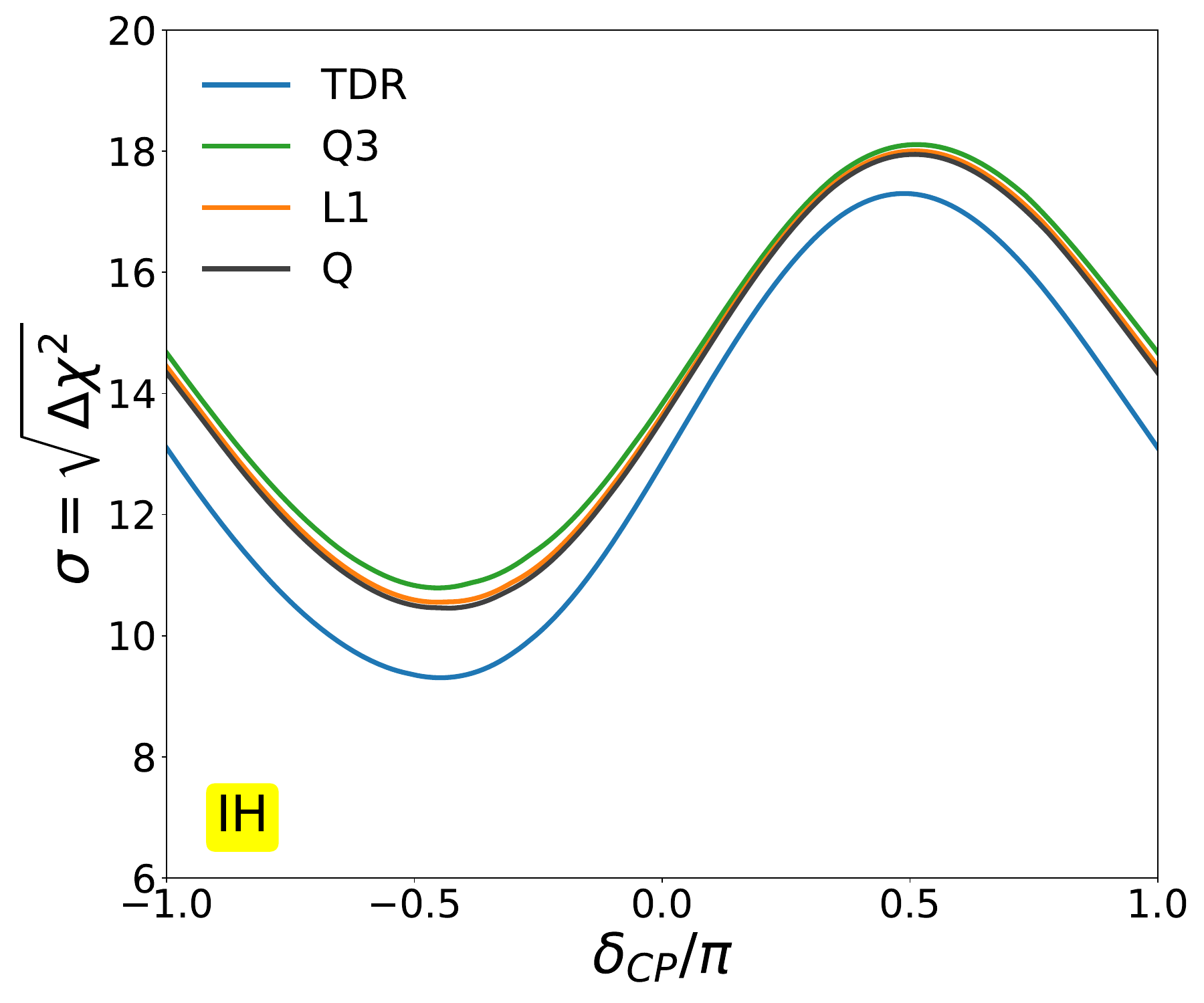} 
    \caption{Significance of neutrino mass ordering determination with the true values of $\delta_{CP}$.}
    \label{fig:mh-sens}
\end{figure}

The determination of the neutrino mass ordering in oscillation experiments primarily relies on the matter effects experienced by neutrinos as they propagate through Earth.\@ These matter effects modify neutrino oscillations differently for the normal and inverted mass orderings~\cite{mo}. With its 1300 km long baseline, DUNE is well-positioned to determine the neutrino mass ordering with high significance~\cite{DUNE:2020ypp}.
To evaluate this capability, we quantify the mass ordering discovery potential using the approach described below.
  \begin{equation}
  \Delta{\chi^2} =
  \begin{cases}
          \chi^2_{IH} - \chi^2_{NH} ~ \text{(for true NH)},\\
      \chi^2_{NH} - \chi^2_{IH} ~ \text{(for true IH)}\,.
  \end{cases}
  \end{equation}
Figure~\ref{fig:mh-sens} presents the significance with which the neutrino mass ordering can be determined as a function of the true values of $\delta_{CP}$, assuming a total runtime of seven years (3.5 years in $\nu$-mode and 3.5 years in $\bar{\nu}$-mode). 
The results are shown for four scenarios: TDR, Q, Q3, and L1, in both the normal and inverted hierarchies. The characteristic shape of the sensitivity curves arises due to the interplay between matter effects and CPV, leading to a near degeneracy at $\delta_{CP} = \pi/2$ ($\delta_{CP} = -\pi/2$) for true NH (IH). It can be seen from the figure, DUNE is capable of resolving the neutrino mass ordering with a minimum significance of 5$\sigma$ across the entire $\delta_{CP}$ range for both NH and IH. Additionally, the Q, Q3, and L1 calorimetry scenarios consistently demonstrate higher sensitivity than the TDR scenario. Furthermore, Q and L1 yield nearly identical results and remain competitive with Q3 in both mass orderings.

\begin{figure}[htb]
    \centering
    \includegraphics[width=0.7\linewidth]{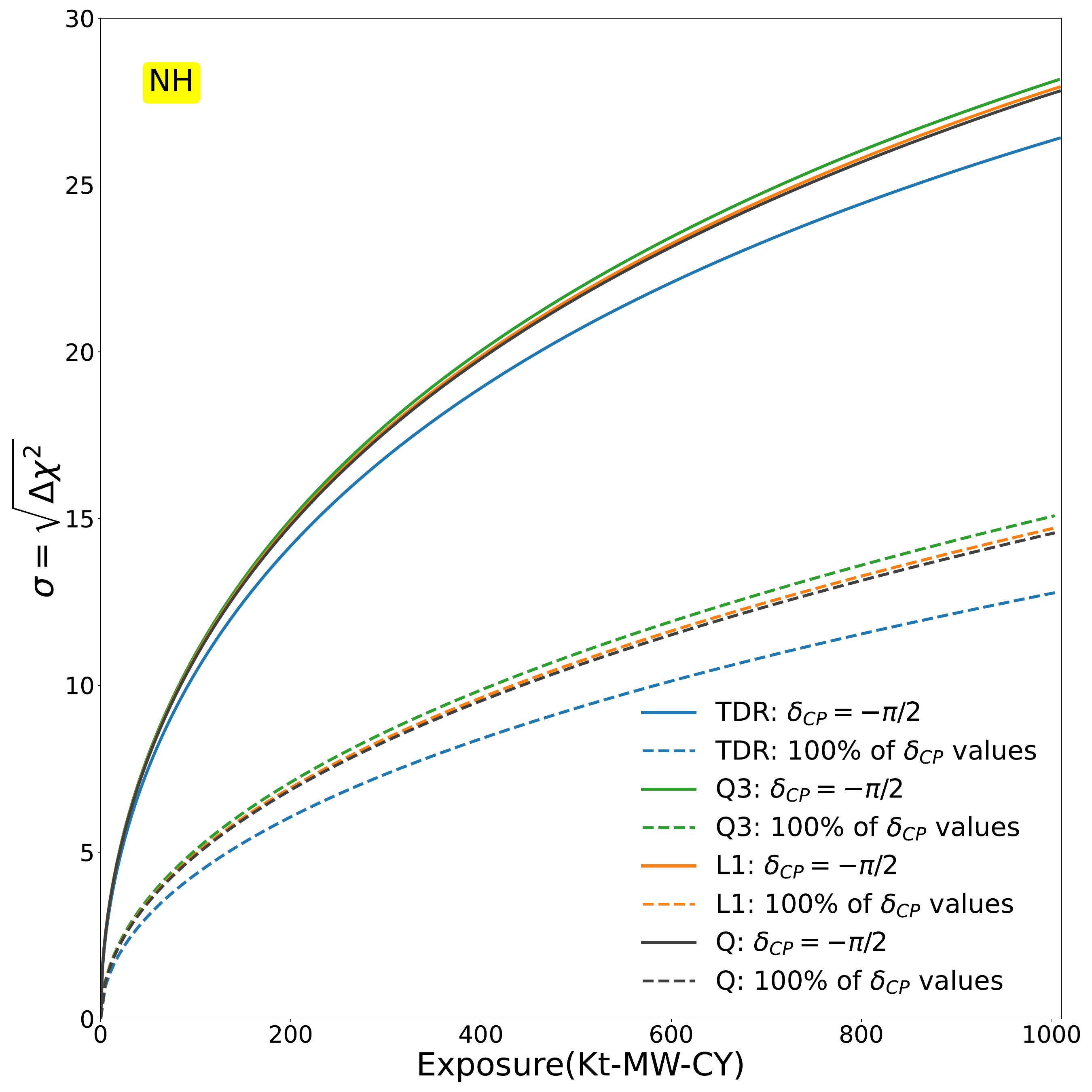} \quad
     \includegraphics[width=0.7\linewidth]{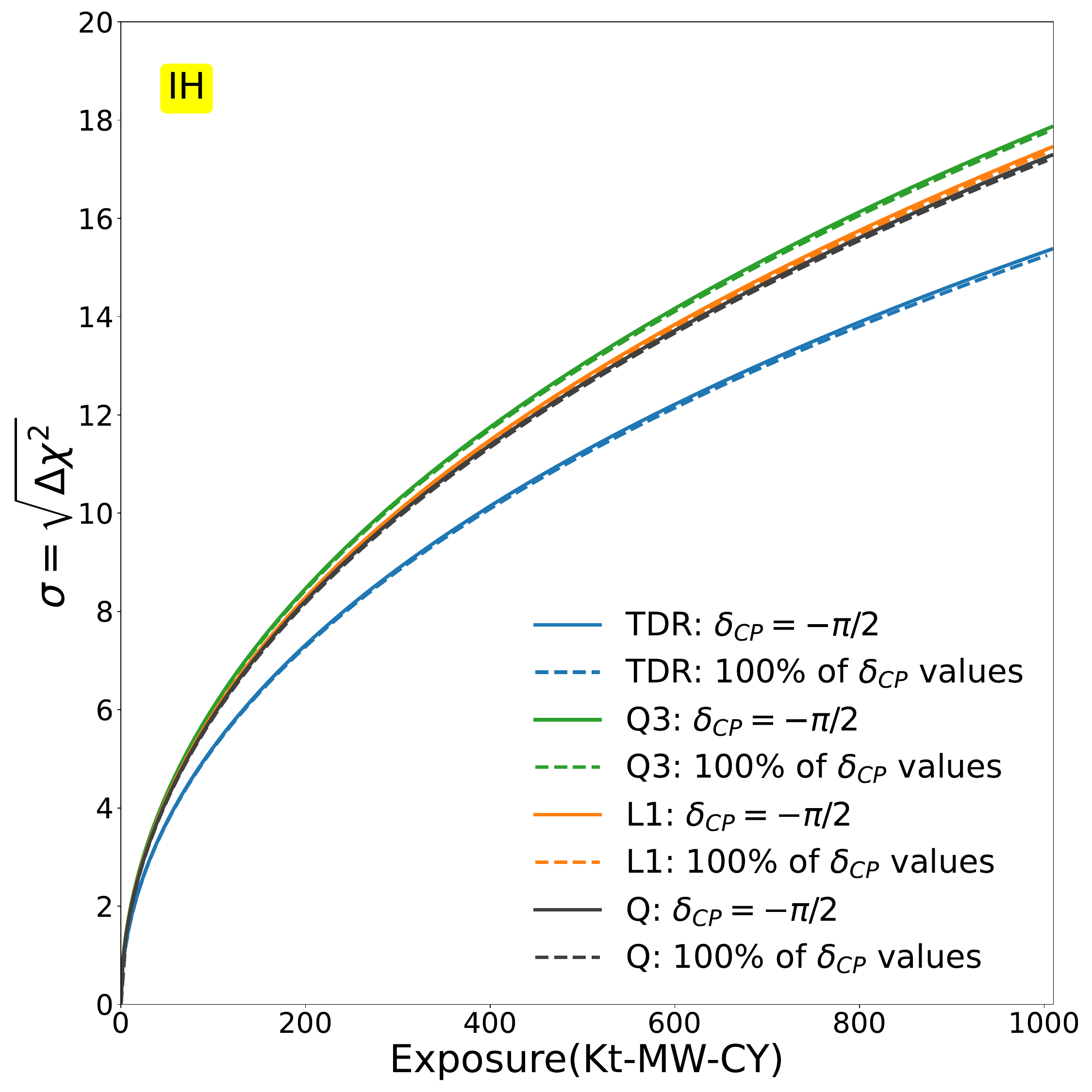} 
    \caption{Significance of neutrino mass ordering determination as a function of exposure (Kt-MW-CY). The solid curves represent the sensitivity for $\delta_{CP}=-\pi/2$, while the dashed curves correspond to the results for 100\% of possible true $\delta_{CP}$ values. }
    \label{fig:mhsens-exp}
\end{figure}

The significance of determining the neutrino mass ordering is presented in figure~\ref{fig:mhsens-exp} for NH (left panel) and IH (right panel), in four scenarios: TDR, Q, Q3, and L1, with exposure in Kt-MW-CY. 
The dashed curves represent the sensitivity for 100\% of $\delta_{CP}$ values, while the solid curves correspond to $\delta_{CP} = -\pi/2$. A 15$\sigma$ discovery can be achieved for $\delta_{CP}=-\pi/2$ in NH (IH) within $4.7 ~(19.9)$ years using TDR, $4.3~ (15.3)$ years with Q, $4.2 ~(15)$ years with L1, and $4.1 ~(14.2)$ years with Q3. For $100\%$ of the values $\delta_{CP}$, a significance of $10\sigma$ is reached in NH (IH) after $12.1 ~(8.1)$ years with TDR, $9.2~ (6.3)$ years with Q, $9 ~(6.2)$ years with L1, and $8.5~ (6)$ years with Q3.
The observed differences  in NH and IH sensitivity arise due to the lowest sensitivity occurring at $\delta_{CP}=-\pi/2$ for NH and at $\delta_{CP}=\pi/2$ for IH. As a result, the sensitivity curves for $100\%$ of $\delta_{CP}$ values closely follow those for $\delta_{CP}=-\pi/2$ in IH. Among all the scenarios, TDR consistently exhibits the lowest sensitivity.

\subsection{Octant Sensitivity}
\begin{figure}[hbt!]
    \centering
    \includegraphics[width=0.65\linewidth]{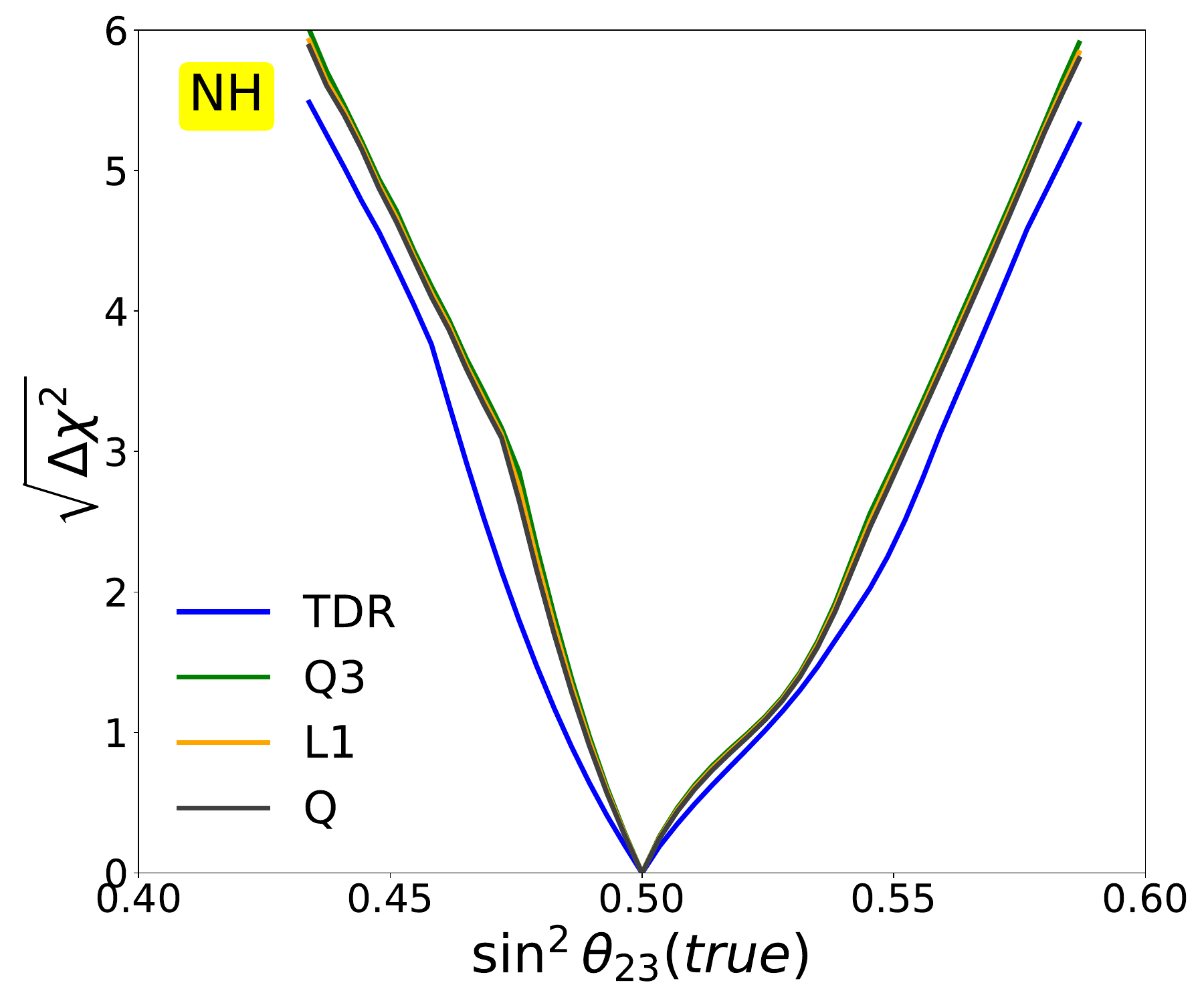} \quad
     \includegraphics[width=0.65\linewidth]{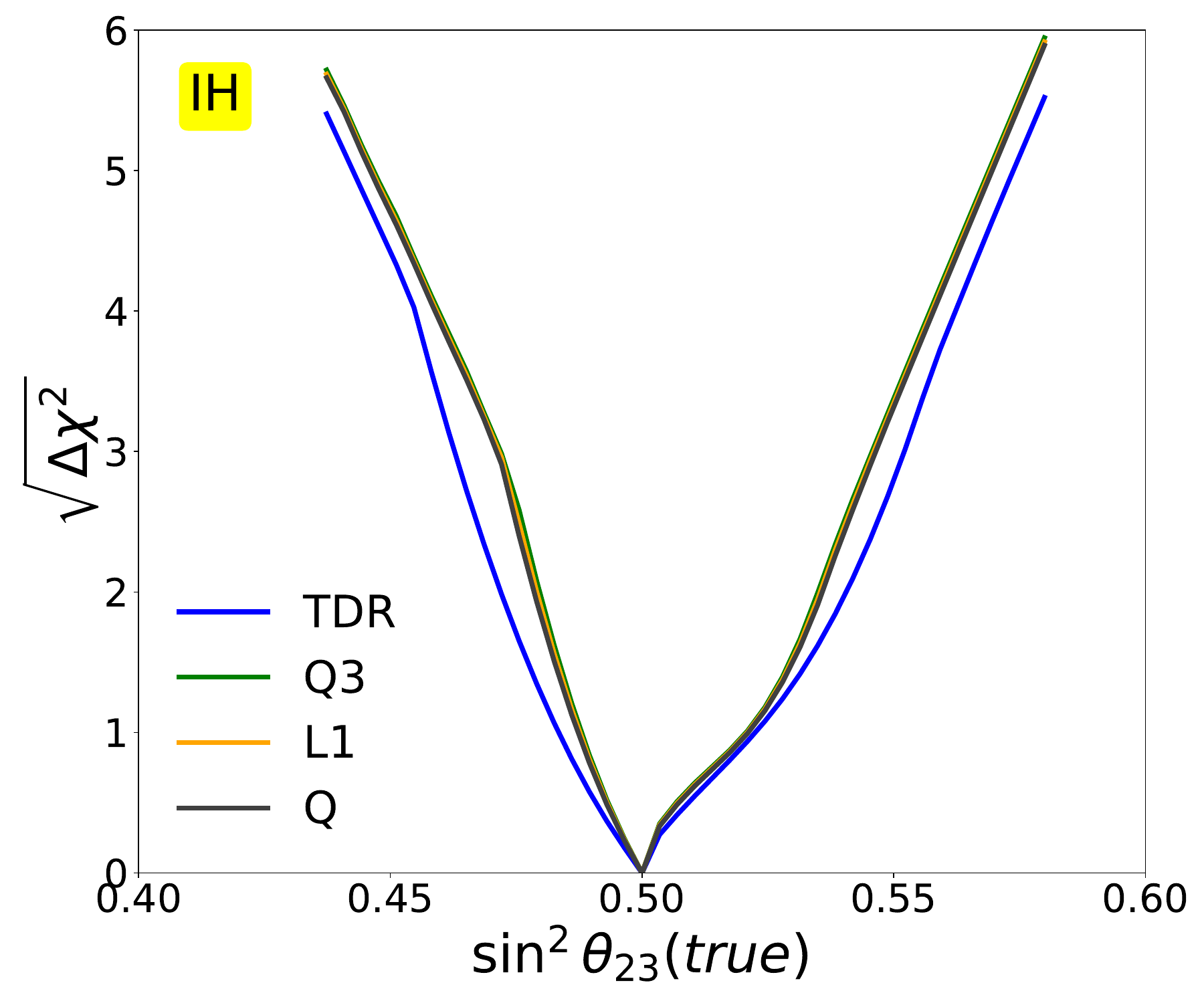} 
    \caption{Significance of the determination of the $\theta_{23}$ octant as a function of the true values of $\sin^2 \theta_{23}$.}
    \label{fig:oct-sens}
\end{figure}

In DUNE, the appearance channel ($\nu_\mu \to \nu_e$) is primarily sensitive to $\sin^2 \theta_{23}$, while the disappearance channel ($\nu_\mu \to \nu_\mu$) depends on $\sin^2 2\theta_{23}$. By combining both channels, DUNE can effectively probe the octant of $\theta_{23}$~\cite{DUNE:2020ypp}. The sensitivity to the octant is quantified using the $\Delta{\chi^2}$ metric, defined as:
\begin{eqnarray}
      \Delta{\chi^2} &=& \chi^2(\pi/2 - \theta_{23}^{true}) - \chi^2(\theta_{23}^{true}).
  \end{eqnarray}
Figure~\ref{fig:oct-sens} illustrates the significance of determining the octant of $\theta_{23}$ as a function of the true values of $\sin^2 \theta_{23}$, assuming a seven-year total runtime (3.5 years in $\nu$-mode and 3.5 years in $\bar{\nu}$-mode) across four different scenarios, with the left panel corresponding to NH and the right panel to IH.
 The chosen range of $\sin^2 \theta_{23}$ is based on the 3$\sigma$ allowed region from Table~\ref{tab1}. These sensitivity curves are obtained after minimizing over the full range of $\dcp^{true} \in [-\pi, \pi]$. The results indicate that the sensitivity to the octant of $\theta_{23}$ is comparable for Q, L1, and Q3, while TDR consistently exhibits lower sensitivity. Additionally, for both NH and IH, the lower octant shows better sensitivity compared to the upper octant.

\section{Conclusion}

In this study, we explored the impact of improved energy resolutions beyond the TDR approach, considering charge based methods such as the approach proposed by A. Friedland et al., the advanced charge calorimetry method, and a simple light calorimetry technique proposed by X. Ning et al. We evaluated their effects on the precise determination of key unknowns in neutrino oscillation physics, including CP violation, mass ordering, and the octant of $\theta_{23}$. Our analysis was conducted using both a fixed exposure of 336 Kt-MW-CY and varying exposures at DUNE. The results demonstrate significant improvements in the measurement of these parameters with enhanced energy resolution, as compared to the standard TDR approach. Notably, the performance of the simple light calorimetry method is particularly promising. The key conclusions from this study regarding the measurement of these unknowns are summarized as follows.

\noindent

\textbf{CP Violation:} Based on the CPV sensitivity results discussed in Section~\ref{results}, we find that charge calorimetry provides the highest sensitivity among the four scenarios. Light calorimetry exhibits a slight advantage over A. Friedland et al.'s method near $\dcp=\pm\pi/2$, independent of the mass ordering. As given in Table~\ref{tab:milestone}, the discovery of CPV is expected to be achieved earlier for IH than NH. A $3\sigma$ or higher sensitivity for $75\%$ of $\dcp$ values can be obtained after approximately $12.4$ years ($12$ years) for NH (IH) using charge calorimetry. In comparison, A. Friedland et al.'s method and light calorimetry require $13$ years ($12.7$ years) and $12.8$ years ($12.5$ years), respectively. While both methods yield similar performance, light calorimetry performs slightly better than A. Friedland et al.'s method. Similarly, a $5\sigma$ or higher sensitivity for $50\%$ of $\dcp$ values is achieved after about $9.4$ years ($8.8$ years) in NH (IH) using charge calorimetry, with light calorimetry outperforming A. Friedland et al.'s method in both mass orderings. Overall, charge, light, and A. Friedland et al.'s calorimetry approaches demonstrate superior sensitivity compared to the TDR, with light calorimetry outperforming A. Friedland et al.'s method.

\begin{table*}[ht]
\centering
%
\begin{tabular}{|c|c|c|c|c|}
\hline
\multirow{5}{*}{Physics Milestone} & \multicolumn{4}{c|}{Exposure in Years} \\
\cline{2-5}
&&&& \\
 &  ~~~~~~~~\underline{TDR}~~~~~~~~ &  ~~~~~~~\underline{A. Friedland et.al.}~~~~~~~ &  ~~~~~~~\underline{Light}~~~~~~~ &  ~~~~~~~\underline{Charge}~~~~~~~  \\
 &  NH (IH) &  NH (IH) &  NH (IH) &  NH (IH) \\
 &&&& \\
\hline
 3$\sigma$ CP violation  & 2.3 (1.6) & 1.8 (1.3) & 1.7 (1.3) & 1.6 (1.2)  \\
 ($\dcp=-\pi/2$)  &  &  &  &  \\
 \hline
 3$\sigma$ CP violation  & 3.6 (3.1) & 3.1 (2.9) &  3 (2.8) & 2.6 (2.7) \\
 (50$\%$ of $\dcp$ values)  &  &  &  &  \\
 \hline
 3$\sigma$ CP violation  & 15 (13.9) & 13 (12.7) & 12.8 (12.5) & 12.4 (12)  \\
 (75$\%$ of $\dcp$ values)  &  &  &  &  \\
 \hline
 5$\sigma$ CP violation  & 7 (4.9) & 5.3 (4)  & 5.2 (3.9)  & 4.7 (3.8) \\
 ($\dcp=-\pi/2$)  &  &  &  &  \\
 \hline
 5$\sigma$ CP violation  & 11.2 (10.3) & 9.5 (9.3) &  9.4 (9.1) & 9.4 (8.8)  \\
 (50$\%$ of $\dcp$ values)  &  &  &  &  \\
\hline
 $\dcp$  Resolution of 10$^\circ$  & 6 (6.3)  & 5.5 (5.9) & 5.4 (5.8)  &  5.3 (5.6)\\
 ($\dcp=0$)  &  &  &  &  \\
 \hline
 $\dcp$  Resolution of 20$^\circ$  & 5.4 (7.2) & 3.8 (4.7) & 3.6 (4.6) &  3.3 (4.2)\\
 ($\dcp=-\pi/2$)  &  &  &  &  \\
 \hline
 5$\sigma$ Mass Ordering  & 2.8 (1.9) & 2.1 (1.5) & 2.1 (1.5)  & 2.0 (1.4) \\
 (100$\%$ of $\dcp$ values)  &  &  &  &  \\
 \hline
 10$\sigma$ Mass Ordering  & 12.1 (8.1) & 9.2 (6.3) & 9 (6.2) & 8.5 (6) \\
 (100$\%$ of $\dcp$ values)  &  &  &  &  \\
\hline
\end{tabular}
\caption{The exposure in runtime, equally split between neutrino and antineutrino modes, needed to achieve the selected physics milestones with TDR, A. Friedland et.al. (Q), Light calorimetry (L1), and Charge calorimetry (Q3).} \label{tab:milestone}
\end{table*}

\noindent

\textbf{CP Phase Resolution:}
The resolution for all four scenarios is poorest near the maximally CP-violating phase, $\dcp=-\pi/2$, and best near the CP-conserving phase, $\dcp=0$. In general, NH provides better resolution than IH. With 20 years of exposure, equally divided between neutrino and antineutrino modes, the achieved resolutions for TDR, A. Friedland et al., Light, and Charge calorimetries are 6.5$^\circ$ (6.5$^\circ$), 6.2$^\circ$ (6.1$^\circ$), 6.2$^\circ$ (6.1$^\circ$), and 6.1$^\circ$ (6.1$^\circ$), respectively, in NH (IH) at $\dcp=0$. Light, A. Friedland et al., and Q3 Charge calorimetries exhibit nearly identical performance and remain highly competitive with each other at large exposures, regardless of mass ordering. As given in Table~\ref{tab:milestone}, a resolution of 10$^\circ$ at $\dcp=0$ can be achieved approximately eight months earlier with Charge calorimetry, seven months earlier with Light calorimetry, and six months earlier with A. Friedland et al. compared to TDR, irrespective of the mass ordering. Similarly, a resolution of 20$^\circ$ at $\dcp=-\pi/2$ can be reached nearly 2 years earlier in NH and 3 years earlier in IH compared to TDR, with Light and A. Friedland et al. calorimetries yielding nearly identical results and remaining competitive with Charge calorimetry.

\textbf{Mass Ordering:}
A 5$\sigma$ significance for determining the neutrino mass ordering for 100\% of $\dcp$ values can be achieved in approximately 2 (1.4) years with Charge calorimetry, 2.1 (1.5) years with Light calorimetry, 2.1 (1.5) years with A. Friedland et al. calorimetry, and 2.8 (1.9) years with TDR in NH (IH). Thus, Charge calorimetry reaches this milestone ninemonths months earlier than TDR for NH and six months earlier for IH. Light and A. Friedland et al. calorimetries exhibit nearly identical performance, independent of the true mass ordering. A 10$\sigma$ significance for mass ordering discovery, covering 100\% of $\dcp$ values, can be attained after approximately 12.1 (8.1) years with TDR, 9.2 (6.3) years with A. Friedland et al., 9 (6.2) years with Light, and 8.5 (6) years with Charge in NH (IH). With a fixed exposure of 336 Kt-MW-CY, a minimum 5$\sigma$ sensitivity is achievable for all values of $\dcp$, regardless of the true mass ordering.
Light and A. Friedland et al. calorimetries perform competitively, with Light slightly outperforming A. Friedland et al. in certain cases. All three calorimetry approaches significantly outperform TDR, demonstrating their effectiveness in enhancing mass ordering sensitivity.

\textbf{Octant of $\mathbf{\theta_{23}}$:} The sensitivity to the octant of $\theta_{23}$ is significantly improved with Charge, Light, and A. Friedland et al. calorimetries compared to the TDR configuration. The statistical significance for determining the octant is consistently higher when $\theta_{23}$ lies in the lower octant than in the higher octant. Furthermore, all three advanced calorimetry methods Charge, Light, and A. Friedland et al. exhibit comparable performance in resolving the octant, whereas TDR demonstrates the weakest sensitivity.

In conclusion, the precise measurement of key neutrino oscillation parameters at DUNE is essential for testing the validity of the standard three flavor neutrino framework. Enhanced energy resolution from advanced calorimetry techniques, such as A. Friedland et al., charge calorimetry, and light calorimetry, significantly improves the discovery potential for CP violation and mass ordering compared to the traditional TDR approach. Among these methods, light calorimetry stands out for its simplicity in reconstructing neutrino energy while providing results that are both complementary to and competitive with charge calorimetry. These improvements collectively strengthen DUNE's capability to deliver groundbreaking insights into the nature of neutrinos and their role in the universe.


\acknowledgments
We thank Xuyang Ning, Wei Shi, Chao Zhang, Ciro Riccio and Jay Hyun Jo for reading our manuscript, for providing feedback and helpful discussions. We sincerely acknowledge Biswaranjan Behera for his invaluable idea, continuous support, and significant contributions to the preparation of this manuscript. JR would like to thank OSHEC (Odisha State Higher Education Council) for the financial support through Mukhyamantri Research Innovation(MRI) for extramural research funding-2023. Please note that this work was conducted independently by the authors and does not reflect the views or represent the DUNE Collaboration. 




\bibliographystyle{JHEP}
\bibliography{biblio.bib}

\end{document}